\documentclass[aps,onecolumn,epsfig,prb]{revtex4}
\usepackage{graphicx}

\begin{document}

\title{Noise and the Measurement Process for a Circular Josephson Array Qubit}

\author{Joachim Sj\"ostrand and Anders Karlhede}
\affiliation{Department of Physics, Stockholm University,
Albanova University Center,SE-10691 Stockholm, Sweden}

\date{\today}


\begin{abstract}
\noindent
We discuss a charge qubit consisting of a circular array of Josephson 
junctions. The  two-level system we 
consider couples the two charge states through a higher order 
tunneling process thus making it possible to achieve a long relaxation 
time. Using the spin-boson Hamiltonian, we  
estimate decoherence due to ohmic as well as $1/f$ noise. 
We simulate the quantum mechanical measurement process by studing the 
density matrix of the qubit and a capacitively coupled single-electron 
transistor that measures the charge.
\end{abstract}

\maketitle

\vskip0.1pc

\section{Introduction}

\noindent
In recent years much effort has been spent on the search for  
quantum two-level systems, qubits, that can be coherently controlled long 
enough for a sequence of controlled unitary operations to be performed 
on them. The ultimate goal is to build a quantum computer out of these qubits. 
Proposals for qubits based on a variety of
physical systems exist, each with its pros and cons. Here we consider  
solid state charge qubits based on 
Josephson junctions (JJ) arrays. 
These have the advantage of being relatively easy to manipulate and 
the prospects for large scale manufacturing are comparatively good. 
However, they suffer from severe decoherence effects.  The single 
Cooper-pair box (SCB) is the simplest proposal for a charge based JJ 
qubit.\cite{SSH97} For the SCB, superposition of charge states 
 was observed by  Bouchiat {\it et al.}\cite{Bou98} and coherent 
 evolution was demonstrated by Nakamura \textit{et al}.\cite{Nak99} 

A generalisation of the SCB, which we call the circular 
array (CA), was introduced by Sch\"ollmann \textit{et al}.\cite{AK02} This circuit
consists of an array of tunable JJs in a circular geometry. 
The CA is similar to the SCB and many results can be taken over {\it mutatis mutandis}. The main 
difference is that the two charge states of the 
qubit are coupled through a higher order tunneling process. Turning 
the coupling off
then allows the tunneling rate to be made very small -- leading to  slow 
relaxation, and a long time to perform the 
measurement. This is the key element of the quantum sample and hold 
(QUASH) measurement strategy.\cite{AK02}

In this article we perform a more detailed study of the circular array. 
In particular,  we consider the effect of voltage fluctuations in the 
circuit (ohmic noise) as well as $1/f$ noise, believed to be caused by 
background charge fluctuations, and calculate the 
relaxation and dephasing times for these types of noise -- extending 
the previous treatment.\cite{AK02} We also study the 
measurement of the qubit's charge by a single-electron transistor 
(SET) coupled capacitively to the CA. This is done by numerically  
determining the time development of the density matrix following the 
treatment of Makhlin {\it et al.}\cite{MSS01} for the 
SCB.

\section{The circular array}

\begin{figure}[tbp]
\begin{center}
\includegraphics[angle=0,width=0.7\textwidth]{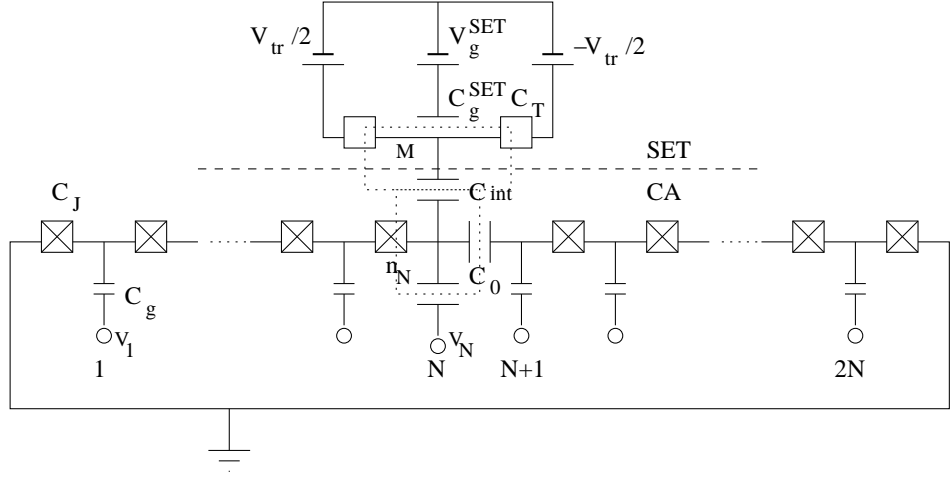}
\caption{The circular array with a SET connected to island $N$. The box 
symbols (without cross) stand for normal junctions. Island $N$ on 
the CA and the island of the SET is marked with dotted boxes.}
\label{CA_circ}
\end{center}
\end{figure}

\noindent
The circular array consists of two arrays with $N$ identical JJs each -- 
these 
arrays are  connected in series and separated by a capacitor $C_0$, 
thus forming a circular geometry. Each JJ, which  is 
a small SQUID, has capacitance $C_J$ and a Josephson energy 
$E_J=E_J(\Phi(t))$ which can be tuned by altering the magnetic flux 
$\Phi$  through the SQUID loop.
The lead connecting the two arrays is grounded to allow  
charge to tunnel in and out of the circuit. There are  
$2N$ small islands, $i=1,2,\ldots2N$, each 
characterized by the  number of excess Cooper pairs $n_i$ and the phase of 
the superconducting order parameter $\phi_i$; these  are quantum 
mechanically conjugate variables: $[\phi_i,n_j] = i \delta_{ij}$. Each island 
charge is externally controlled by a gate voltage $V_i(t)$, 
applied via a small capacitor $C_g$. Fig.~\ref{CA_circ} shows the 
circular array together with the SET that measures the charge on 
one of the islands next to $C_{0}$.

A qubit should have two states separated by a large gap, $\delta E$, 
from higher energy states and be weekly coupled to the environment to 
avoid rapid decoherence. The CA fulfills this if  $C_{0}\sim C_{J}\gg C_{g}$.  
The energy scales present in the system is the charging energy for a 
Cooper pair $E_C \equiv (2e)^2/2C_J \sim \delta E$, the Josephson energy $E_J$, the 
superconducting gap $\Delta$ and the temperature $k_B T$. In order to 
avoid quasiparticles in the system at low temperatures, the qubit is 
constructed so that $\Delta$ is the largest energy in 
the problem.   Furthermore, we choose the qubit to be in the charge regime $E_C \gg 
E_J$,  and impose $E_{C}\gg k_{B}T$ 
to avoid thermal excitation of higher energy charge states, thus:
\begin{equation}
\label{E_ineq}
\Delta \gg E_C \gg E_J,k_B T \ .
\end{equation}

The Hamiltonian of the CA is
\begin{equation}
\label{H}
H = H_{C}+H_{J}=\frac{1}{2} \sum_{i,j=1}^{2N} Q_i C_{ij}^{-1} Q_j - E_J 
\sum_{\langle ij \rangle } \cos(\phi_i - \phi_j ) \ ,
\end{equation}
where $Q_i=2e(n_i -n_{g,i})$ is the effective 
charge on island $i$ -- here $n_{g,i} =C_g V_i/2e$ is the gate charge 
on the island. 
$C_{ij}$ is the capacitance matrix -- its nonzero elements are: 
$C_{N,N}=C_{N+1,N+1}=C_0 + C_J + C_g$, $C_{i,i}=2C_J+C_g$,  
$C_{N+1,N}=C_{N,N+1}=-C_0$ and  $C_{i+1,i}=C_{i,i+1}=-C_J$, where 
$i\neq N,N+1$. 
The matrix is symmetric and $C_{ij}=C_{2N+1-i,2N+1-j}$; 
the inverse matrix $C^{-1}$
has the same symmetries.   The sum over the 
Josephson terms in (\ref{H}) is taken over all pairs of islands 
connected by tunnel junctions.

Since we are studying a charge qubit it is convenient to write the 
Hamiltonian in the charge basis $|\textbf{n}\rangle = 
|n_1 n_2 ...n_{2N}\rangle$. The charging energy term simply becomes
$H_C= E_C C_J \sum_{\textbf{n}} (\textbf{n} - 
\textbf{n}_g)^t C^{-1} (\textbf{n} - \textbf{n}_g) |\textbf{n}\rangle 
\langle \textbf{n}|$ where $\textbf{n}_g=(n_{g,1},n_{g,2}, \dots, 
n_{g,2N})$, and using that  $|n_i\rangle = 
\int_0^{2\pi} d \phi_i e^{-in_i\phi_i}|\phi_i \rangle$, which holds since $n_i$ 
and $\phi_i$ are conjugate, the Josephson term becomes
\begin{equation}
\label{Hj}
H_J = -\frac{E_J}{2} \sum_{\textbf{n},\langle ij \rangle}  \prod_{k 
\neq i,j} |n_k \rangle \Big( |n_i+1\rangle |n_j-1 \rangle + 
|n_i-1\rangle |n_j+1 \rangle \Big ) \langle \textbf{n}|
\end{equation}
and the total Hamiltonian is $H = H_C(\textbf{V}) + H_J(\Phi)$,
where we have indicated the dependence on the external control 
parameters $\textbf{V} = (V_1,V_2, \dots, V_{2N})$ and $\Phi$.

\begin{figure}[tbh]
\begin{center}
\includegraphics[angle=270,width=0.47\textwidth]{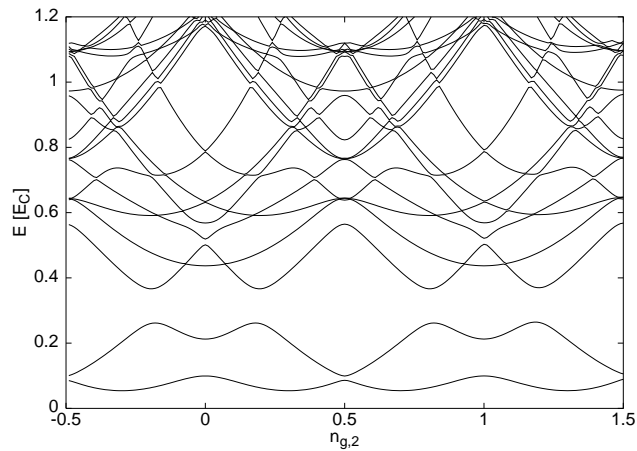}
\caption{The energy levels for the $N=2$ CA as functions of 
$n_{g,2}$. The other parameters are $E_J=0.2E_C$, $n_{g,3}=0.5$ and 
$C_J=C_0=100C_g$. (Charge states with $-2\leq n_i \leq 2$ are included 
in the diagonalization.)}
\label{CA_energies}
\end{center}
\end{figure}

The two-level system -- {\it ie} the qubit -- we consider consists 
of states $\mid\uparrow \rangle$, 
$\mid\downarrow \rangle$  with one excess 
Cooper pair on either of the islands $N,N+1$ neighbouring $C_0$: 
$\mid\uparrow \rangle \equiv |0\dots 10\dots0\rangle$ and 
$\mid\downarrow \rangle \equiv |0 \dots01 \dots0\rangle$, where the 
ones are for island $N$ and $N+1$ respectively. For $E_J=0$ these two states 
are degenerate if $n_{g,N}=n_{g,N+1}=\frac{1}{2}$ and 
$n_{g,i}=0$ for $i\neq N, N+1$. If, in addition, $C_0 \sim C_J \gg 
C_g$, then the energy gap 
to higher charge states is $\delta E \sim E_C$. 
Restricting ourselves to a finite charge space, the Hamiltonian $H$
can be  diagonalised numerically. In 
Fig.~\ref{CA_energies} we show the energy spectrum for the 
$N=2$ CA as a function of $n_{g,2}$. The other parameters are 
$n_{g,3}=1/2$, $C_0=C_J=100 C_g$, $E_J=0.2 E_C$ and $-2 \leq n_i \leq 
2$ (this restriction gives a negligible error). 
For $n_{g,2} \approx 1/2$, the two lowest energy levels (which 
are linear combinations of the states $\mid\uparrow \rangle$, $\mid\downarrow 
\rangle$ with an excess Cooper pair on island $N$ or $N+1$ respectively) 
form a  two-level system with a large gap to the higher energy states. 

We conclude that when $E_C \gg E_J$, $C_0\sim C_J \gg C_g$ and 
$\textbf{n}_g \approx (0,\dots, \frac{1}{2},\frac{1}{2}, \dots, 0)$, 
it is a good approximation to restrict the Hilbert space of the 
Hamiltonian in Eq.~(\ref{H}) to the states $\mid\uparrow \rangle$ 
and $\mid\downarrow \rangle$ defined above. We write the Hamiltonian of this 
two-level system in spin-1/2 notation
\begin{equation}
\label{H2lvl}
H_{ctrl}(t) = -\frac{1}{2} B_z(\delta V(t)) \sigma_z -\frac{1}{2} 
B_x(\Phi(t)) \sigma_x \ ,
\end{equation}
where $\sigma_i$ are the Pauli matrices in the basis 
$(\mid\uparrow \rangle$, $\mid\downarrow \rangle)$. This Hamiltonian controls 
the qubit -- unitary operations can be 
performed on the qubit by tuning $B_z$ and $B_x$
via the external parameters $\delta V= V_{N+1}-V_{N}$ and $\Phi$.  Writing 
$\textbf{n}_g=(0,\dots,\frac{1+\delta n_g}{2},\frac{1-\delta 
n_g}{2},\dots,0)$ where $\delta n_g = C_g \delta V/2e$, we find $B_z = \langle\downarrow\mid H_C\mid\downarrow \rangle - \langle\uparrow |H_C|\uparrow \rangle =
2A_N E_C \delta n_g$, where $A_N = C_J ( C_{NN}^{-1} - C_{N,N+1}^{-1})$. $A_N$ 
can be calculated numerically for given capacitance matrix, however,  
we can also perform an expansion in 
$C_g/C_0$ which is valid as long as $N$ is not too large. 
Using Cramer's rule for the elements in $C^{-1}$, we  
express $A_N$ in terms of cofactors and expand.  Assuming, for 
simplicity, $C_J=C_0$, this gives
\begin{equation}
\label{BZ}
B_z = \frac{2N}{2N+1} \Big[ 1 - 
\frac{C_g}{C_0}\frac{N+1}{6} + {\cal{O}} ((C_{g}/C_{0})^{2})\Big] 
\times E_C \delta n_g \ \ \ .
\end{equation}
(For $N=2$ and to leading order in $C_g$, this reproduces a previous result.\cite{AK02})
$B_x$ gives the cotunneling rate of a Cooper pair from island $N$ to 
$N+1$ via the $2N$ junctions and and its leading contribution is obtained by $(2N)^{th}$ order perturbation theory, hence $B_x \sim (E_J/E_C)^{2N}$. The exact numerical factor is not very 
illuminating -- it can however be determined for not too large $N$. For 
the $N=2$ CA, with parameters as in Fig.~\ref{CA_energies}, we have 
$B_z \approx \frac{4}{5} \Big[ 1 - \frac{1}{2}\frac{C_g}{C_0} \Big] 
E_C \delta n_g = 0.796 E_C \delta n_g$ and $B_x = 
\frac{125}{12}(E_J/E_C)^4 E_C = 0.0167 E_C$ \cite{AK02}. (A numerical 
diagonalisation gives $B_z = 0.7960 E_C \delta n^g$ and $B_x 
= 0.0132 E_C$ -- in reasonable agreement with the expansions taking 
into account that the ignored terms are of order
${\cal O} ((C_{g}/C_{0})^{2})\sim 10^{-4}$ and ${\cal O} ((E_{J}/E_{C})^{6})\sim 10^{-4}$ respectively.) 

Diagonalizing (\ref{H2lvl}) gives the 
eigenvalues  $\pm\frac{1}{2} \Delta E$ where $\Delta E = \sqrt{B_x^2+B_z^2}$, 
with corresponding eigenvectors $|+\rangle  = 
\cos\frac{\theta}{2} |\uparrow \rangle + 
\sin\frac{\theta}{2} |\downarrow \rangle, \nonumber$ and 
$|-\rangle  =  -\sin\frac{\theta}{2} |\uparrow \rangle + 
\cos\frac{\theta}{2} |\downarrow \rangle$, 
where $\theta = \arctan (B_x/B_z)$. If  $\tau$ denotes the  Pauli matrices 
in the energy  eigenbasis, then  $H_{ctrl}$ can be written 
in the compact form
\begin{equation}
\label{Hfinal}
H_{ctrl} = -\frac{1}{2} \Delta E \tau_z \ . 
\end{equation}



\section{Electromagnetic noise}

\noindent
The practical usefulness of a circuit like the CA as a qubit is 
ultimately limited by the coupling to external degrees of freedom. These lead to
decoherence of the qubit state and hence loss of quantum information to the environment.  The generic behaviour of the evolution 
of the qubit depends  on the strength of the coupling  and one 
identifies two regimes:
The ``Hamiltonian-dominated'', where the coupling to 
the environment is weak enough for the time evolution of the qubit 
to be governed by the qubit Hamiltonian $H_{ctrl}$, and the 
``environment-dominated'', where the coupling to the environment is 
so strong that it determines the 
dynamics of the qubit. In this 
article we consider only the Hamiltonian-dominated regime. Note, 
however,  that even if the coupling to the environment is weak under 
normal operation of the qubit, it becomes environment-dominated if 
$H_{ctrl}\approx 0$, which may happen during the qubit manipulations. 

In the Hamiltonian-dominated regime the evolution of the qubit is 
conveniently described in the energy eigenbasis ($|-\rangle $, 
$|+\rangle $). The interaction with 
the environment leads to a decay of the off-diagonal elements in the 
qubit's density matrix with a charecteristic time $\tau_{\varphi}$,
the dephasing time,
\begin{equation}
\label{tp_1}
\langle \tau_\pm (t) \rangle = \langle \tau_\pm (0) \rangle e^{\mp i 
\Delta E t} e^{-t/\tau_\varphi } \ ;
\end{equation}
whereas the diagonal elements of the density matrix decay to their thermal 
equilibrium values with a characteristic time $\tau_{relax}$, the 
relaxation time,
\begin{equation}
\label{tr_1}
\langle \tau_z (t) \rangle = \tau_z(\infty) + ( \tau_z(0)- 
\tau_z(\infty)) e^{-t/\tau_{relax}} \ ,
\end{equation}
where the thermal equilibrium value is 
$\tau_z(\infty)=\tanh(\Delta E/2 k_BT)$. 

A Josephson junction charge qubit is sensitive to various electromagnetic  
fluctuations in the circuit; we follow standard practice and model 
these with the  ``spin-boson'' model \cite{CL83} with an Ohmic 
spectrum. In addition to the noise caused by these fluctuations one 
observes $1/f$ noise, which is believed to be due to background 
charge fluctuations in the substrate. Following Shnirman \textit{et al},\cite{MSS02} we model this phenomenologically using again the spin-boson model but now with a $1/f$-spectrum.

The spin-boson model describing the qubit interacting with the environment 
has one independent bath for each island in the CA: 
\begin{equation}
\label{Hsb}
H_{SB} = H_{ctrl} + \sigma_z \sum_{i=1}^{2N} X_i + \sum_{i=1}^{2N} 
H_B^i  \ .
\end{equation}
Here, $H_{ctrl}$ is the qubit Hamiltonian (\ref{H2lvl}),
$H_B^i = \sum_{i,a} \Big ( \frac{p_{ai}^2}{2m_{ai}} + 
\frac{m_{ai}\omega_{ai}x_{ai}^2}{2} \Big )$ is a bath of harmonic  
oscillators with coordinates $x_{ai}$, momenta $p_{ai}$, masses $m_{ai}$ 
and frequencies $\omega_{ai}$.  The baths lead to voltage fluctuations 
$X_i \equiv \sum_a C_{ai} x_{ai}$ coupling to  $\sigma_{z}$.  ($C_{ai}$ is the 
strength of the coupling between the qubit and the a'th oscillator in 
bath $i$.)

The effect of the environment is completely 
characterized by a spectral function, which for the spin-boson 
model in Eq.~(\ref{Hsb}) has the form
\begin{equation}
\label{Jw}
J(\omega) = \frac{\pi}{2} \sum_{a,i} 
\frac{C_{ai}^2}{m_{ai}\omega_{ai}} \delta(\omega - \omega_{ai}) = 
\frac{\pi}{2} \sum_i \alpha^i_s \hbar \omega_{00}^{1-s} \omega^s 
\Theta(\omega_c-\omega) \ .
\end{equation}
To obtain the second equality one assumes that $J(\omega)$ can be 
written as a power of $\omega$ up to some cut-off frequency 
$\omega_{c}$ which is assumed to be large compared to all other 
frequencies in the problem. The parameter $s$ 
reflects the qualitative nature of the environment, and $\alpha_s 
\equiv \sum_i \alpha_s^i$ is a dimensionless measure of the strength 
of the coupling. To maintain $\alpha_s$ dimensionless for all $s$, 
an additional frequency scale $\omega_{00}$ enters for $s\neq 1$. 

\begin{figure}[tbh]
\begin{center}
\includegraphics[angle=270,width=0.5\textwidth]{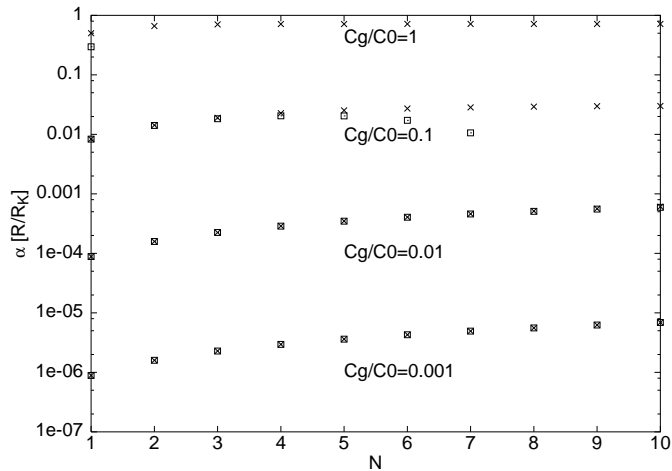}
\caption{$\alpha_1$ as a function of $N$ for different $C_g/C_0$. The 
numerical result is shown as crosses and the analytic expansion as 
squares. $\alpha_{1}$ is given in units of $R/R_K$ and the scale on 
the $y$-axis is logarithmic; $R_i=R$ 
for all islands, and $C_J=C_0$.}
\label{aN}
\end{center}
\end{figure}

We model the voltage fluctuations $\delta V_i$ on island $i$ by 
adding an impedance $Z_i(\omega)$ in series with $V_i$, see, {\it eg}, 
Ingold and Nazarov\cite{IN92} or Makhlin {\it et al.}\cite{MSS99} This 
impedance then has Johnson-Nyquist fluctuations $\delta V_i$ between 
its terminals that in the spin-boson formalism correspond to the 
spectral function  $J_i(\omega) = \omega Re \Big[ Z_{it}(\omega)\Big]$, 
where  $Z_{it}(\omega)$ is the total impedance between the terminals 
of $Z_i$. From Fig.~\ref{CA_circ} one finds $Z_{it}(\omega) = 
\Big[ i\omega C_{\Sigma i} + Z_i^{-1}(\omega) \Big]^{-1}$ where
$C_{\Sigma i}^{-1}=C_g^{-1}+(s_iC_0)^{-1}$, with 
$s_i=i^{-1}+(2N+1-i)^{-1}$ (assuming  $C_J=C_0 \gg C_g$). Following 
standard practice, we assume that the noise is 
purely resistive, $Z_i(\omega) = R_i$, and if furthermore $R_i \ll 1/\omega 
C_{\Sigma i}$ (which holds for realistic $R_i$ and $C_g$ since $C_{\Sigma,i}<C_g$), 
we obtain $J_i(\omega)=\omega R_i$, 
which is linear in $\omega$ and hence corresponds to $s=1$ in 
Eq.~(\ref{Jw}). Using this we can obtain the total spectral 
function $J(\omega)$ for the circuit. From Eq.~(\ref{Hsb}) we identify 
$X=-\frac{1}{2}\delta B_z$, where $\delta B_z = \Big[ 
\langle\downarrow |H_C|\downarrow\rangle - \langle\uparrow\mid 
H_C\mid\uparrow \rangle \Big] \Big|_{n^g_i=\frac{C_g}{2e}\delta 
V_i}$. Simplifying this expression yields $X =-e\frac{C_g}{C_J} 
\sum_{i=1}^{2N} A_i \delta V_i$, with $A_i \equiv C_J \Big ( 
C_{N+1,i}^{-1} - C_{N,i}^{-1} \Big )$. The fluctuations $\delta V_i$ 
are expressed in the oscillator coordinates, $\delta V_i = \sum_a 
b_{ia} x_{ia}$, with the spectral function in the form of 
Eq.~(\ref{Jw}), $J_i(\omega)=\frac{\pi}{2} \sum_a 
\frac{b_{ia}^2}{m_{ia} \omega_{ia}} \delta(\omega-\omega_{ia})= 
\omega R_i$. This gives $X=-e\frac{C_g}{C_J} \sum_{i=1}^{2N} A_i 
\sum_a b_{ia} x_{ia}$ and hence, $J(\omega)= \omega \sum_i \Big[ 
-e\frac{C_g}{C_J}A_i \Big]^2 R_i$. Comparing this to  
Eq.~(\ref{Jw}), with $s=1$, we find
\begin{equation}
\label{av_exact}
\alpha_1 = 4 \sum_i A_i^2 \cdot \frac{R_i}{R_K} \Big( \frac{C_g}{C_J} 
\Big)^2,
\end{equation}
where $R_K = h/e^2 = 25.8 \ k\Omega$ is the quantum of 
resistance. Assuming the islands to be nearly identical, it is reasonable 
that $R_i$ are approximately the same for all islands. 
Setting $R_i=R$, $i=1,\ldots2N$,  we are 
left with the factor $\sum_i A_i^2$, which we can calculate  
numerically for given capacitance matrix or as an  expansion in $C_g/C_0$.
The expansion is analogous to the one for $B_{z}$ in Eq.~(\ref{BZ}) -- 
assuming for simplicity $C_{J}=C_{0}$ we obtain
\begin{equation}
\label{av}
\alpha_1 = \frac{4N(N+1)}{3(2N+1)} \Bigg[ 1 - 
\frac{2C_g}{15C_0}(N^2+N+3) + {\mathcal{O}}((C_g/C_0)^2) \Bigg] \times 
\frac{R}{R_K} \Bigg( \frac{C_g}{C_0} \Bigg )^2.
\end{equation}
(For $N=2$ and to leading order in $C_g$, Eq.~(\ref{av}) gives $\alpha_1 = 
\frac{8}{5} \frac{R}{R_K} \Big ( \frac{C_g}{C_J} \Big )^2$ as derived previous.\cite{AK02}) Fig.~\ref{aN} shows $\alpha_1$ as a function of $N$ for different $C_g/C_0$. The numerical result is 
shown as crosses, and the expansion to order $(C_{g}/C_{0})^{3}$ as 
squares.
For 
$C_g/C_0=10^{-2}$ and $10^{-3}$, the expansion agrees very well with
the numerical result (at least when $N \leq 10$). For $C_g/C_0=0.1$ and 
$1$, the expansion becomes negative at $N\geq 8$ and $N\geq 
2$ respectively. Therefore no squares are seen in this region. In a 
typical circuit we may have $R \sim 50 \ \Omega$, yielding $R/R_K 
\sim 10^{-3}$. For $C_g \ll C_0$, $\alpha_1 \ll 1$ for all realistic 
$N$.

The characteristic times $\tau_{relax}$ and $\tau_{\varphi}$ can  
be calculated within the spin-boson model using
perturbation theory or path integral methods.\cite{Leg87} For 
$s=1$ they are 
\begin{eqnarray}
\label{t_rel_h}
\Gamma_{relax} \equiv \tau_{relax}^{-1} & = & \frac{1}{\hbar^2} \sin^2 
\theta \cdot S_X(\Delta E/ \hbar)
 = \pi \alpha_1 
 \frac{\Delta E}{\hbar} \coth \frac{\Delta E}{2 k_B T} \sin^2 \theta \ , \\
\label{t_dep_h}
\Gamma_{\varphi} \equiv \tau_{\varphi}^{-1} & = & 
\frac{1}{2}\Gamma_{relax} + \frac{1}{\hbar^2} \cdot S_X(0)
 = \frac{1}{2}\Gamma_{relax} + \pi \alpha_1 \theta \frac{2k_B 
T}{\hbar}  \cos^2 
\theta \ ,
\end{eqnarray}
where 
\begin{equation}
\label{Sw}
S_X(\omega) \equiv \langle [X(t), 
X(t')]_+\rangle_\omega = 2 \hbar J(\omega) \coth \frac{\hbar 
\omega}{2 k_B T}
\end{equation}
is the Fourier transform of the symmetrized correlation function at 
thermal equilibrium. These results can be  applied to  the CA 
simply by substituting  $\alpha_1$ from Eq.~(\ref{av_exact}). One 
defines the pure dephasing rate $\Gamma_{\varphi}^\ast$, as 
$\Gamma_\varphi = \frac{1}{2}\Gamma_{relax}+\cos^2 \theta 
\Gamma_{\varphi}^\ast$. The Hamiltonian-dominated regime is 
realized when $\Delta E \gg \hbar \Gamma_{\varphi}^\ast$ (at least 
for $s=0,1$ \cite{Wei92}). For ohmic damping, 
$\Gamma_{\varphi}^\ast=2\pi \alpha_1 k_B T/\hbar$, and the 
condition becomes $\Delta E \gg \alpha_1 
k_B T$.

Noise with a power spectrum proportional to the inverse of the frequency 
is observed in many  physical systems. 
This  $1/f$ noise is seen also in JJ circuits \cite{Nak01}, where it is 
 believed to be caused by background charge 
fluctuations.\cite{Zor96} Here we follow Shnirman \textit{et al},\cite{MSS02} 
and model this noise with the 
spin-boson model. For $s=0$ and $\omega \ll k_B T /\hbar$, (\ref{Sw}) 
gives the wanted  power spectrum: $S_X(\omega) = E_{1/f}^2 / 
\omega$ with $E_{1/f}^2 = 2 \pi \hbar \alpha_0 \omega_{00} k_B T$ 
(here $T$ is an adjustable parameter). 
Sub-ohmic environments ({\it ie} those for which $0\leq s<1$) have not been 
much studied as it was believed that they rapidly localized the system in one of 
the $\sigma_{z}$-eigenstates for 
any strength of the  damping. However, it has been realized that this 
is true only for large damping, whereas for weak damping the system 
behaves coherently\cite{KH96}. The times $\tau_{relax}$ and $\tau_{\varphi}$ have been 
calculated for $1/f$ noise for the single bath spin-boson model in 
the Hamiltonian-dominated regime\cite{MSS02}. The relaxation rate is
\begin{equation}
\Gamma_{relax} =   \frac{E_{1/f}^2}{\hbar \Delta E} \sin^2 \theta \ .
\end{equation}
The dephasing rates are only known for $\theta=0,\pi/2$; they are
\begin{equation}
\label{tf_1of_0}
\Gamma_\varphi = 
\frac{E_{1/f}}{\hbar}\sqrt{\frac{1}{\pi}\ln \frac{E_{1/f}}{\hbar 
\omega_{ir}}}  \ \ \ \ \ \ \ \ \ {\rm for} \ \ \ \theta=0
\end{equation}
and
\begin{equation}
\label{tf_1of_pi2}
\Gamma_\varphi =  \frac{E_{1/f}^2}{\hbar \Delta E} \frac{1}{2\pi} \ln 
\frac{E_{1/f}^2}{\hbar \omega_{ir}\Delta E} \ \ \ \ \ \ \ \ \ {\rm 
for} \ \ \  \theta=\pi/2 \ ,
\end{equation}
both with logarithmic accuracy in $E_{1/f}$ \cite{MSS02}. Here, $\omega_{ir}$ is an 
infrared cut-off frequency which can be experimentally determined. To 
determine the times we need to determine $E_{1/f}$.\footnote{The 
relaxation rate is in fact given by  Eq.~(\ref{t_rel_h}) with $S_X(\omega) = E_{1/f}^2 / 
\omega$, whereas Eq.~(\ref{t_dep_h}) cannot be used for the dephasing 
rates since $S_X(0) \to \infty$ for $s=0$.}




\section{The quantum measurement of charge}

\noindent
The state of the qubit is inferred by performing a quantum 
measurement of the charge using a SET  coupled capacitively to 
island $N$  of the circular array, see  Fig.~\ref{CA_circ}. We 
simulate this quantum measurement by  studying 
the time development of the density matrix describing  the CA and the 
SET -- we follow closely Makhlin \textit{et al},\cite{MSS01} where the corresponding
problem is treated for the SCB. 

The SET is a circuit with one normal island surrounded by two 
junctions connected to normal electrodes and a capacitor $C_g^{SET}$, 
see Fig.~\ref{CA_circ}. During manipulations of the qubit, no current 
flows through  the SET -- the SET is turned off -- this is 
achieved by setting the transport voltage to zero, $V_{tr}=0$, and 
tuning the gate voltage $V_g^{SET}$ away from a degeneracy point so 
that the Coulomb blockade suppresses the tunneling through the SET. 
When the manipulations of the qubit are done and one wants to read out 
the result, then the Coulomb blockade is turned off by tuning the gate  
voltage to a degeneracy point and a transport voltage is turned on leading 
to a tunneling current through the SET. This current depends on the 
state  of the qubit through the charge on the capacitor $C_{int}$.
This leads to a measurement of the charge of the qubit.

The  density matrix for the CA and the SET can be written as $\hat{\rho} \equiv 
\hat{\rho}(i,i',M,M',m,m')(t)$ after one has traced out the microscopic 
degrees of freedom in the left and right electrodes and in the 
island of the SET. 
Here, $i$ labels the state of the qubit, $M$ is the 
number of (excess) electrons on the island in the SET and $m$ is 
the number of electrons that have tunneled through the SET. It is 
possible to derive a master equation for $\hat{\rho}$ as 
an expansion in the SET tunneling terms.\cite{SS94} For low
temperature and small $V_{tr}$, only 
transitions between two adjacent charge states of the SET need to be 
taken into account (these states are assumed to be $M=0,1$ below). If one 
furthermore assumes that the tunneling is instantaneous then one obtains a 
set of simple equations for the diagonal matrix elements 
$\hat{\rho}^{ij}_M(m,t) \equiv \hat{\rho}(i,j;M,M;m,m)(t)$. 
In terms of the Fourier transformed  quantity  
$\hat{\rho}^{ij}_M(k,t) = \sum_m e^{-ikm} \hat{\rho}^{ij}_M(m,t)$,  
the final form is a system of eight coupled differential equations 
(for each $k$)
\begin{equation}
\label{Me}
\hbar \frac{d}{dt} {\hat{\rho}_0 \choose \hat{\rho}_1} + 
{i[H_{ctrl},\hat{\rho}_0] \choose i[H_{ctrl} + \delta 
H_{int},\hat{\rho}_1] } = {-\check{\Gamma}_L \quad 
e^{-ik}\check{\Gamma}_R \choose \check{\Gamma}_L \quad 
-\check{\Gamma}_R} {\hat{\rho}_0 \choose \hat{\rho}_1} \ ,
\end{equation}
where $\hat{\rho}_M$ is short for the $2 \times 2$ matrix 
$\hat{\rho}^{ij}_M(k,t)$. $H_{ctrl}$ is the qubit Hamiltonian with a 
renormalised capacitance matrix ($C_{N,N} \to C_{N,N}+C_{int}$) due 
to the presence of the SET, $\delta H_{int}=E_{int} \sigma_z$ is the 
coupling energy where $E_{int}$ is 
determined by the capacitances. The tunneling rates $\Gamma_{L/R}$ are 
\begin{eqnarray}
\check{\Gamma}_L \hat{\rho}_0 \equiv \Gamma_L \hat{\rho}_0 + \pi 
\alpha_L [\delta H_{int},\hat{\rho}_0]_+ \ , \nonumber \\
\check{\Gamma}_R \hat{\rho}_1 \equiv \Gamma_R \hat{\rho}_1 - \pi 
\alpha_R [\delta H_{int},\hat{\rho}_1]_+ \ ,
\end{eqnarray}
where $\alpha_{L/R}$ and $\Gamma_{L/R}$ is the tunneling conductance 
and the tunneling rate for the left/right junction of the SET. 

We want to study the current through the SET as a function of time --  
this is obtained from
\begin{equation}
\label{Pmt}
P(m,t) \equiv \sum_{i,M} \hat{\rho}(i,i,M,M,m,m)(t) \ ,
\end{equation}
which can be interpreted as the probability that $m$ electrons have 
tunneled through the SET during time $t$. To compute $P(m,t)$, we 
solve the differential equation (\ref{Me}) with suitable initial 
conditions. We assume that the qubit and the SET are disentangled
initially, $\hat{\rho}(t=0) = \hat{\rho}_0^{qb} \otimes 
\hat{\rho}_0^{SET}$, and that the qubit is prepared in some state $| \psi 
\rangle = \alpha \mid\uparrow\rangle + \beta \mid\downarrow\rangle$, 
 $\hat{\rho}_0^{qb} = |\psi \rangle \langle \psi|$. At time 
$t=0$, no electrons have tunneled through the SET, thus 
$P(m,0)=\delta_{m,0}$ and $P(k,0)=1$ for the Fourier transform. From 
the definition (\ref{Pmt}), we find $P(k,0)=Tr(\hat{\rho}_0)+Tr(\hat{\rho}_1)$ and we may choose $(\hat{\rho}_0,\hat{\rho}_1) |_{t=0} = (\mathbf{0},\hat{\rho}_0^{qb})$. 

\begin{figure}[tbp]
\begin{center}
\includegraphics[angle=0,width=0.49\textwidth]{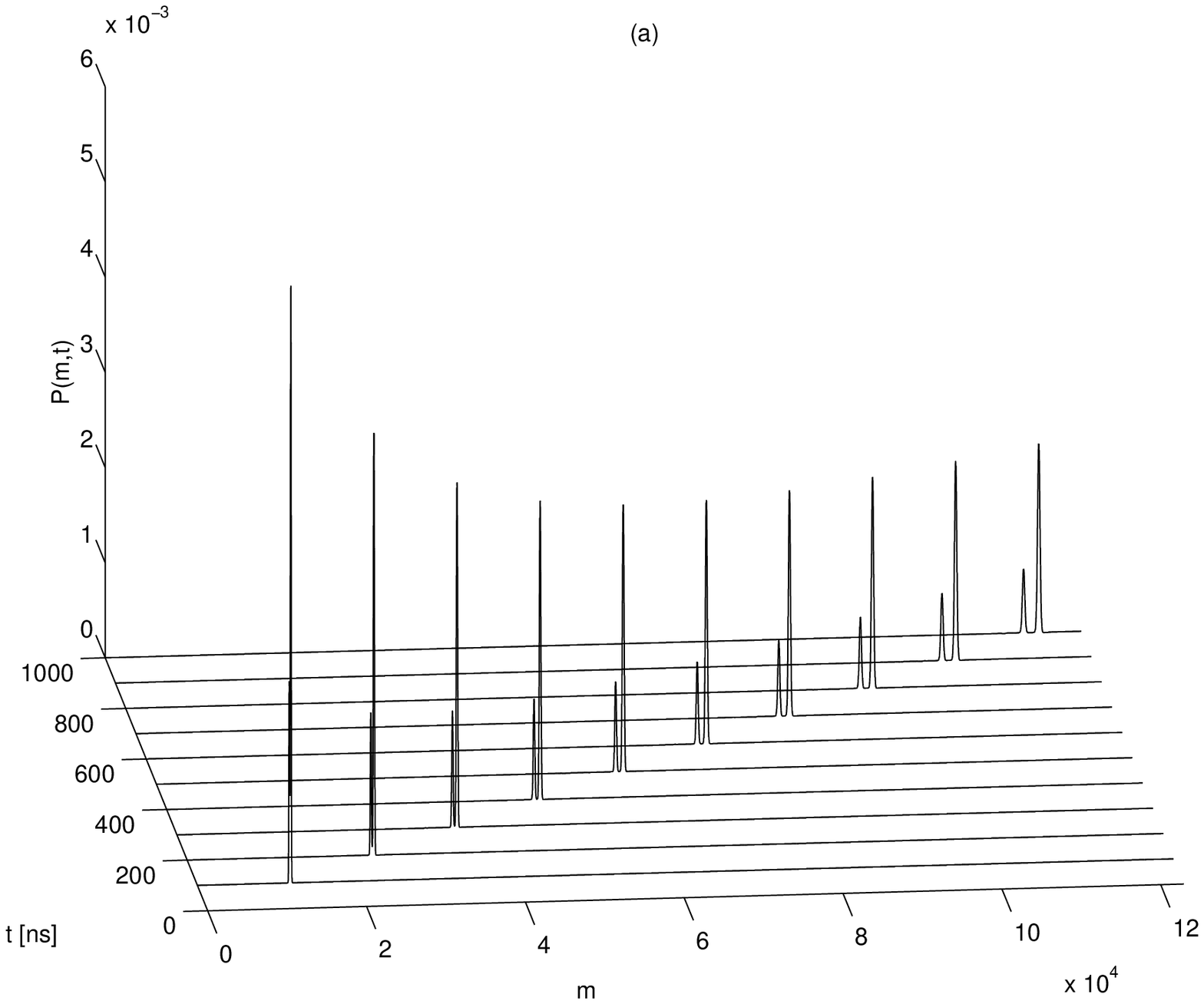}
\includegraphics[angle=0,width=0.49\textwidth]{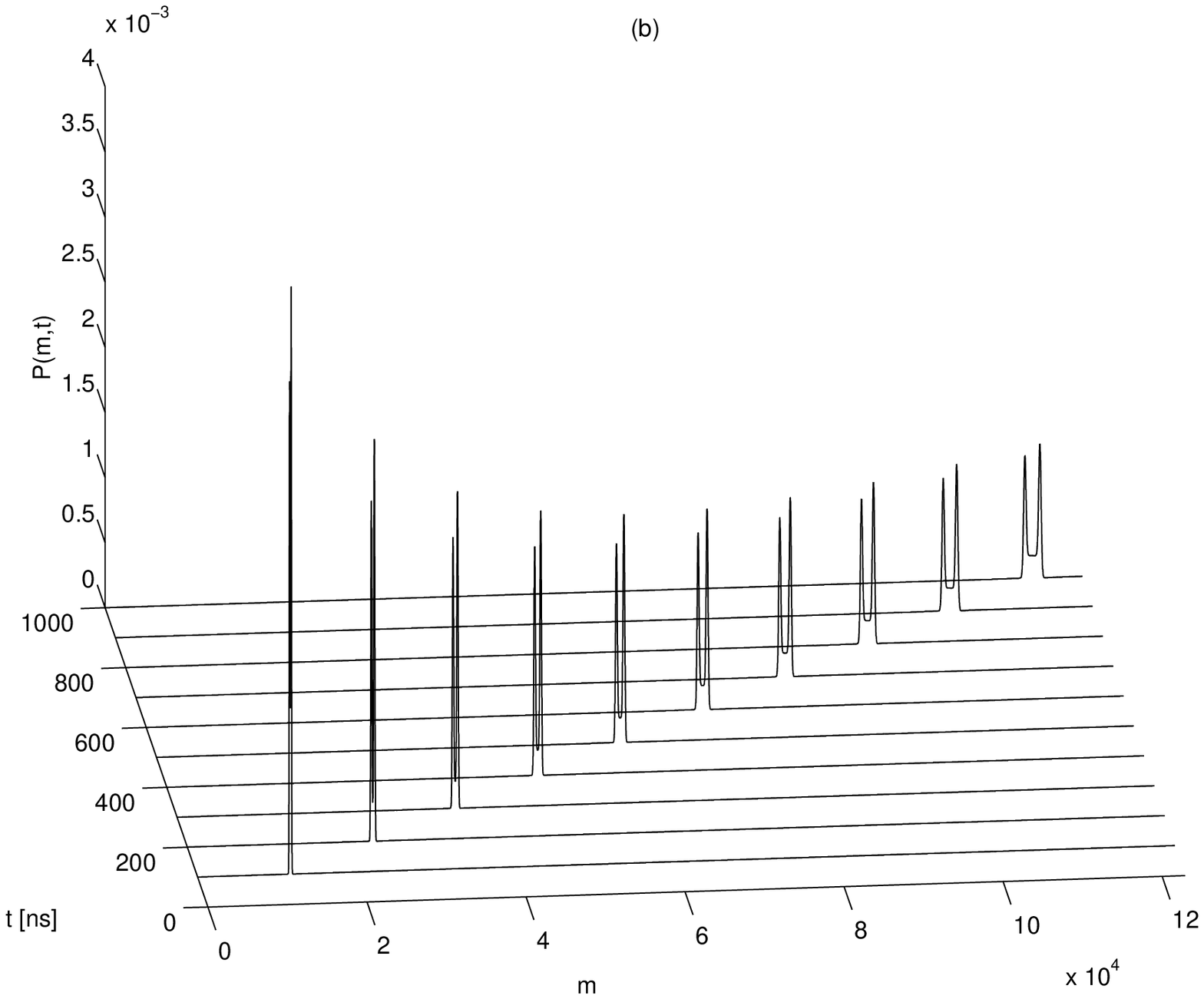}
\caption{$P(m,t)$ as a function of $m$ for ten different times $t$. In (a), 
$B_x=0$, $B_z=9.0 \ \mu$eV and $|\alpha|^2=0.75$,  whereas in (b), 
$B_x=1.3 \ \mu$eV, $B_z=15 \ \mu$eV and $|\alpha|^2=0.50$.}
\label{meas_p_1}
\end{center}
\end{figure}

To calculate $P(m,t)$ for the $N=2$ CA we need  
values for the parameters $E_{int}$, $\alpha_L$, $\alpha_R$, 
$\Gamma_L$ and $\Gamma_R$ for the SET, and $B_z$ and $B_x$ for the 
qubit. We choose 
$C_g=C_{int}=C_g^{SET}=\frac{1}{3}C_T = \frac{1}{100}C_J=\frac{1}{100}C_0 = 0.032$ fF, $\alpha_L = \alpha_R = 0.03$, $\Gamma_L=30\ \mu$eV and $\Gamma_R = 100\ \mu$eV (the qualitative result will not be too sensitive to this choice). Hence 
$E_C = 100 \ \mu$eV, $E_{int} = 
6.1 \ \mu$eV and $B_z = 80 \delta n_g \ \mu$eV. At the start of the 
measurement $B_{x}$ is turned off, $B_x \simeq 0$.  $B_z$ is kept close to 
the degeneracy point, although  $B_z \neq 0$ to avoid  the 
environment-dominated regime.  Fig.~\ref{meas_p_1} 
shows examples of $P(m,t)$.  
In (a) a measurement in the off-state, 
$B_{x}=0$, is shown, the other  parameters are 
$B_z=9.0 \ \mu$eV (corresponding to $\delta n_g 
= 0.11$) and $|\alpha|^2=0.75$. For comparison, 
Fig.~\ref{meas_p_1}(b) shows a mesaurement where $B_{x}$ is on; the 
parameters are in this case $B_x=1.3 \ \mu$eV (which is the maximum value of 
$B_x$ assuming $E_J^{max}=0.2E_C$), $B_z=15 \ \mu$eV 
(corresponding to $\delta n_g = 0.19$) and $|\alpha|^2=0.50$. 
We see that after a while, $P$ develops a two peak 
structure. This is interpreted  as follows. At a given time, the total 
current is a superposition of two different currents flowing in the 
SET, with weights given by the amplitudes, $\alpha$ and 
$\beta$, of the charge states in the qubit state. 
For increasing $t$,  the peaks move towards 
higher $m$ values since  more electrons have tunneled 
through the SET. During this process, the peaks widen, their magnitude 
decreases and the distance (in $m$) between the peaks increases. 
The ratio between the two peak values are $2.96$ in Fig.~\ref{meas_p_1}(a) and 
$1.10-1.20$ (the value decreases with time) in Fig.~\ref{meas_p_1}(b) to be compared to the ratios $|\alpha|^2/|\beta|^2=3$ and $1$ respectively. Fig.~\ref{meas_p_1}(a) corresponds to our proposed measurement 
situation -- the off-state -- and we see that here the peaks are well 
separated after a while and the ratio of their peak values stay close to 
$|\alpha|^2/|\beta|^2$ for a very long time. Thus allowing for a slow accurate measurement 
of the charge. In Fig.~\ref{meas_p_1}(b) on the other hand, the peaks 
are less well separated and the ratio agrees less well with 
$|\alpha|^2/|\beta|^2$. Eventually the valley between the 
peaks will fill out and one broad peak forms due to 
mixing between the charge states -- a plateau indicating this process is clearly seen in (b).
To perform a measurement of the charge, the peaks must have moved 
apart to become well separated and not yet started to merge due to 
mixing. This gives a window in time for measuring the charge -- in 
the off-state in Fig.~\ref{meas_p_1}(a) this window is very large -- 
the lower bound is $\tau_{meas} \sim 500$ ns. The upper bound is not seen, however, we expect it to be considerably larger.

\section{Discussion}
\label{QSE_ex}
\noindent

We here estimate the decoherence times for the circular array. To be 
specific we  assume $C_0=C_J=100C_g$, 
$E_C \simeq 5 E_J^{max} \simeq 100 \mu$eV and $T \simeq 40$ mK $= 3 \ 
\mu$eV and, initially $N=2$ -- higher $N$ are commented on below.

Calculations are performed by tuning $H_{ctrl}(t)$ in time. During this 
process the SET is off, and we are close to the degeneracy point. 
Assuming  $|\delta n_g| \lesssim 0.2$, then
$|B_z| \lesssim 0.16E_C \simeq 16 \ \mu$eV -- in addition we have $|B_x| \lesssim 0.013E_C = 1.3 \ \mu$eV. There is also a lower bound on $\Delta E$, 
since the system is assumed to be in the Hamiltonian-dominated 
regime. The typical time per operation of the qubit is $\tau_{op} \simeq \hbar/\Delta E \sim 10$ ps assuming $\Delta E \simeq 10 \mu$eV.  


For the ohmic noise, we use $R \simeq 50 \ \Omega$, which gives 
$\alpha_1 \simeq 3 \cdot 10^{-7}$. Assuming $\Delta E \lesssim 16 \ \mu$eV 
one finds $\tau_\varphi \sim 100 \ \mu$s and  $\tau_{relax} \gtrsim 100 \ \mu$s.  The Hamiltonian regime is realised when $\Delta E \gg \alpha_{1}k_{B}T \approx 10^{-12}\ $eV. 

For $1/f$ noise, $\tau_\varphi$ is only known for $\theta =0$ and 
$\theta=\pi/2$. We use these two cases to estimate $\tau_{\varphi}$, 
assuming that this gives the correct order of magnitude for general 
$\theta$.  Nakamura \textit{et al.} \cite{Nak01} 
measured the factor $\alpha_{1/f}^{SCB} = (E_{1/f}^{SCB})^2/E_C^2$ 
for the SCB and obtained $\alpha_{1/f}^{SCB} \sim 10^{-6}$. If 
the $1/f$-noise is caused by background charge fluctuations then it is 
reasonable to assume that for the CA $\alpha_{1/f} \simeq 2N 
\alpha_{1/f}^{SCB}$, since the CA has $2N$ islands instead 
of the single island in the SCB. This gives $E_{1/f}^2 = 2\pi \hbar \alpha_0 
\omega_{00} k_B T \simeq 2N E_C^2 \alpha_{1/f}^{SCB}$ and, for the 
parameters above, we find $E_{1/f}^2 \simeq 4\cdot 10^{-6} E_C^2$. 
From Nakamura \textit{et al.} \cite{Nak01} we also take 
$\omega_{ir}\simeq 310$ Hz. For $\theta=0$ we find $\tau_{relax} 
\to \infty$ and $\tau_\varphi \sim 1$ ns if $B_z < 16 \ \mu$eV  and for
$\theta=\pi/2$, we find $\tau_{relax} \sim 10$ ns and $\tau_\varphi 
\sim 5$ ns 
if $B_x < 1.3 \ \mu$eV. This shows that the Hamiltonian-dominated regime is 
realised if $\Delta E \gg 0.5 \ \mu$eV. 

After the calculations, the SET is turned on and  the measurement 
is started by tuning $B_x$ to $0$. The relaxation is now very slow: $\tau_{relax} \to \infty$ when $\theta \to 0$ for both ohmic and $1/f$ noise. This gives ample time to measure $|\alpha|^2$ without using an 
ultra-fast detector, {\it c.f.} Fig.~\ref{meas_p_1}. The dephasing is very rapid, thus the quantum state of the qubit is 
destroyed in a short time but it is still possible to measure the charge. 
This is good enough as a read-out mesurement but not as part of an 
error correction protocol.

For higher $N$, the ohmic decoherence times will decrease somewhat, due to an increase in $\alpha_1$. The same is true for $1/f$ noise, since $E_{1/f} \sim \sqrt{N}$. 

We conclude that it is the $1/f$ noise that limits the operation of 
the CA as a qubit -- it leads to the decoherence time  $\tau_{decoh} \sim 1$ ns. The decoherence due to ohmic noise is much slower. Hence, in practice, the ohmic noise seems to be of little concern. If the typical time for a quantum operation is $\tau_{op} \sim 10$ ps, then the $1/f$ noise  restricts one to $N_{op} \sim 100$ operations, assuming $B_x$ and $B_z$ are restricted to values that realise the Hamiltonian-dominated regime. This is a severe restriction and once again underscores that $1/f$ noise is a serious problem for Josephson junction charge qubits. One possible solution to this might be the echo-technique demonstrated by Nakamura \textit{et al.} \cite{Nak01}.

\begin{center}
\Large{\bf Acknowledgments }\\
\end{center}
We wish to thank Peter \AA gren, Jochen Walter, Silvia Corlevi, 
Janik Kailasvuori and Hans Hansson for useful discussions and 
Sergei Isakov for his 
computer skills. Anders Karlhede was supported by the Swedish Science 
Research Council.\\


\begin{thebibliography}{99}

\bibitem{SSH97} A. Shnirman, G. Sch\"on and Z. Hernon, \textit{Phys. 
Rev. Lett.} \textbf{79}, 2371 (1997).
\bibitem{Nak99} Y. Nakamura, Y. A. Pashkin and J. S. Tsai, Nature 
\textbf{398}, 786 (1999).
\bibitem{Bou98} V. Bouchiat, D. Vion, P. Joyez, D. Esteve and M. H. 
Devoret, \textit{Physica Scripta} vol T76, 165-170 (1998).

\bibitem{AK02} V. Sch\"ollmann, P. \AA gren, D. B. Haviland, T. H. 
Hansson and A. Karlhede, \textit{Phys. Rev. B} \textbf{65}, 
020505(R) (2002).
\bibitem{MSS99} Y. Makhlin, G. Sch\"on and A. Shnirman, 
\textit{Nature} \textbf{386}, 305 (1999).
\bibitem{MSS01} Y. Makhlin, G. Sch\"on and A. Shnirman, \textit{Rev. 
Mod. Phys.}, \textbf{73}, 357 (2001).
\bibitem{CL83} A. O. Caldeira and A. J. Leggett, \textit{Ann. Phys. 
(N.Y.)} \textbf{149}, 374 (1983).
\bibitem{MSS02} A. Shnirman, Y. Makhlin and G. Sch\"on, 
cond-mat/0202518 (2002).
\bibitem{IN92} G.-L. Ingold and Y. Nazarov, \textit{Single charge 
tunneling}, NATO ASI Series, Vol. B 294, edited by H. Grabert and M. 
H. Devoret (Plenum Press, New York, 1992).
\bibitem{Leg87} A. J. Leggett \textit{et al.}, \textit{Rev. Mod. 
Phys.} \textbf{59}, 1 (1987).
\bibitem{Nak01} Y. Nakamura, Yu. A. Pashkin, T. Yamamoto and J. S. 
Tsai, \textit{Phys. Rev. Lett.} \textbf{88}, 047901 (2002).
\bibitem{Wei92} U. Weiss, \textit{Quantum dissipative systems}, World 
Scientific, Singapore (1992).
\bibitem{Zor96} A. B. Zorin \textit{et al.}, \textit{Phys. Rev. B} 
\textbf{53}, 13682 (1996).
\bibitem{KH96} S. Kehrein and A. Mielke, \textit{Phys. Lett. A} 
\textbf{219}, 313 (1996).
\bibitem{SS94} H. Schoeller and G. Sch\"on, \textit{Phys. Rev. B} 
\textbf{50}, 18436 (1994).
\end{thebibliography}
\end{document}